\documentclass[pdftex,twocolumn,epjc3]{svjour3}          
\pdfoutput=1
\RequirePackage[T1]{fontenc}
\RequirePackage{graphicx}
\RequirePackage{mathptmx}      
\RequirePackage{flushend}
\RequirePackage[colorlinks,citecolor=blue,urlcolor=blue,linkcolor=blue]{hyperref}

\journalname{Eur. Phys. J. C}

\begin{document}

\title{Lepton identification at particle flow oriented detector for the future $e^{+}e^{-}$ Higgs factories }


\author{Dan Yu\thanksref{addr1, addr2}
        \and
        Manqi Ruan\thanksref{e2,addr1} 
      \and
      Vincent Boudry\thanksref{addr2}
      \and
      Henri Videau\thanksref{addr2}
}

\thankstext{e2}{e-mail: ruanmq@ihep.ac.cn}

\institute{IHEP, China\label{addr1}
          \and
          LLR, Ecole Polytechnique, France\label{addr2}     
}

\date{Received: date / Accepted: date}

\maketitle

\begin{abstract}

The lepton identification is essential for the physics programs at high-energy frontier, especially for the precise measurement of the Higgs boson. For this purpose, a Toolkit for Multivariate Data Analysis (TMVA) based lepton identification (LICH\footnote{Lepton Identification for Calorimeter with High granularity}) has been developed for detectors using high granularity calorimeters.
Using the conceptual detector geometry for the Circular Electron-Positron Collider (CEPC) and single charged particle samples with energy larger than 2 GeV, LICH identifies electrons/muons with efficiencies higher than 99.5\% and controls the mis-identification rate of hadron to muons/electrons to better than 1\%/0.5\%. Reducing the calorimeter granularity by 1-2 orders of magnitude, the lepton identification performance is stable for particles with E > 2 GeV.
Applied to fully simulated eeH/$\mu\mu$H events, the lepton identification performance is consistent with the single particle case: the efficiency of identifying all the high energy leptons in an event, is 95.5-98.5\%.

\end{abstract}

\section{Introduction}

After the Higgs discovery, the precise determination of the Higgs boson properties becomes the focus of particle physics experiments.
Phenomenological studies show that the physics at TeV scale would be revealed if the Higgs couplings could reach the percent level measurement accuracy\cite{ILCTDR}\cite{peskin}. 

        The LHC is a powerful Higgs factory. 
        However, the precision of Higgs measurements at the LHC is limited by the huge QCD background, the large theoretical and systematical uncertainties. 
        In addition, the Higgs signal at the LHC is usually tagged by the Higgs decay products, making those measurements always model dependent. 
        Therefore, the precision of Higgs couplings at the HL-LHC is typically limited to 5-10\% level depending on theoretical assumptions \cite{atlas}\cite{cms}.

 In terms of Higgs measurements, the electron-positron colliders play a role complementary to the hadron colliders with distinguishable advantages. 
 Many electron-positron Higgs factories have been proposed, including the International Linear Collider (ILC), the Compact LInear Collider (CLIC), the Future e+e- Circular Collider (FCC-ee) and the CEPC \cite{ILCTDR}\cite{clic}\cite{cepcprecdr}. 
 These proposed electron-positron Higgs factories pick and reconstruct Higgs events with an efficiency close to 100\%, and determine the absolute value of the Higgs couplings.
 Compared to the LHC, these facilities have much better accuracy on the Higgs total width measurements and Higgs exotic decay searches, in addition the accuracies of Higgs measurements are dominated by statistic errors.
 For example, the circular electron-positron collider (CEPC) is expected to deliver 1 million Higgs bosons in its Higgs operation, with which the Higgs couplings will be measured to percent or even per mille level accuracy\cite{cepcprecdr}. 

  The lepton identification is essential to the precise Higgs measurements. 
 The Standard Model Higgs boson has roughly 10\% chance to decay into final states with leptons, for example, H$\rightarrow$ WW* $\rightarrow$llvv/lvqq, H$\rightarrow$ZZ*$\rightarrow$llqq, H$\rightarrow \tau\tau$, H$\rightarrow \mu\mu$, etc.
  The SM Higgs also has a branching ratio Br(H$\rightarrow$bb) = 58\%, while the lepton identification provides an important input for the jet flavor tagging and the jet charge measurement. 
  On top of that, the Higgs boson has a significant chance to be generated together with leptons. 
  For example, in the ZH events, the leading Higgs generation process at 240-250 GeV electron-positron collisions, about 7\% of the Higgs bosons are generated together with a pair of leptons ( Br(Z$\rightarrow$ee) and Br(Z$\rightarrow\mu\mu$) = 3.36\% ). 
At the electron-positron collider, ZH events with Z decaying into a pair of leptons is regarded as the golden channel for the HZZ coupling and Higgs mass measurement\cite{mumuh}. 
Furthermore, leptons are intensively used as a trigger signal for the proton colliders to pick up the physics events from the huge QCD backgrounds. 
         The Particle Flow Algorithm (PFA) becomes the paradigm of detector design for the high energy frontier\cite{cmsupg,pfa,cepcprecdr,eepfa}.
         The key idea is to reconstruct every final state particle in the most suited sub-detectors, and reconstruct all the physics objects on top of the final state particles. 
         The PFA oriented detectors have high efficiency in reconstructing physics objects such as leptons, jets, and missing energy. 
         The PFA also significantly improves the jet energy resolution, since the charged particles, which contribute the majority of jet energy, are usually measured with much better accuracies in the trackers than in the calorimeters \cite{Arbor,pfa,cmspfa,jcpfa,henripfa}. 

          To reconstruct every final state particle, the PFA requires excellent separation by employing highly-granular calorimeters.
          In the detector designs of the International Large Detector (ILD) or the Silicon Detector (SiD) \cite{ILCTDR,ildloi}, the total number of readout channels in calorimeters reaches the $10^{8}$ level.
         In addition to cluster separation, detailed spatial, energy and even time information on the shower developments is provided. 
         An accurate interpretation of this recorded information will enhance the physics performance of the full detector \cite{FDmanqi}.

          Using the information recorded in the high granularity calorimeter and the dE/dx information recorded in the tracker, LICH(Lepton Identification in Calorimeter with High granularity), a dedicated lepton identification algorithm for Higgs factories has been developed. Using CEPC conceptual detector geometry \cite{cepcprecdr}(based on ILD) and the Arbor\cite{Arbor} reconstruction package, its performance is tested on single particles and physics events. 
          For the single particles with energy higher than 2 GeV, LICH reaches an efficiency better than 99.5\% in identifying the muons and the electrons, and 98\% for pions. Its performance on physics events (eeH/$\mu\mu$H) and the final efficiency agrees with the efficiency at the single particle level. 

          This paper is organized as follows. The detector geometry and the samples are presented in section 2. In section 3, the discriminant variables measured from charged reconstructed particles are summarized and the algorithm architecture is presented. In section 4, the LICH performance on single particle events is presented. In section 5, the correlations between LICH performance and the calorimeter geometry are explored. In section 6, the LICH performance on ZH events where Z decays into ee or $\mu\mu$ pairs is studied, the results are then compared with that of single particle events. In section 7, the results are summarized and the impact of calorimeter granularity is discussed. 

\section{Detector geometry and sample}

 In this paper, the reference geometry is the CEPC conceptual detector \cite{cepcprecdr}, which is developed from the ILD geometry \cite{ILCTDR}. ILD is a PFA oriented detector meant to be used for centre of mass energies up to 1 TeV. It is equipped with a low material tracking system and a calorimeter systems with extremely high granularity.  
 
 In this CEPC conceptual detector design, the forward region, and the yoke thickness have been adjusted to the CEPC collision environment with respect to the ILD detector.
 The core part of this detector is a large solenoid of 3.5 Tesla. The solenoid system has an inner radius of 3.4 meters and a length of 8.05 meters, inside which both tracker and calorimeter system are installed.  
     The tracking system is composed of a TPC as the main tracker, a vertex system, and the silicon tracking devices. The amount of material in front of the calorimeter is kept to $\sim$ 5\% radiation length. 
 Both ECAL and HCAL use sampling structures and have extremely high granularity. The ECAL uses tungsten as the absorber and silicon for the sensor. 
 In depth, the ECAL is divided into 30 layers and 
 in the transverse direction, each layer is divided into 5 by 5 mm$^2$ cells. 
 The HCAL uses stainless steel absorber and GRPC(Glass Resistive Plate Chamber) sensor layers. It uses  10 by 10 mm$^2$ cells and has 48 layers in total. 

As a Higgs factory, the CEPC will be operated at 240-250 GeV center of mass energy. 
To study the adequate lepton identification performance, we simulated single particle samples (pion+, muon-, and electron-) over an energy range of 1-120 GeV (1, 2, 3, 5, 7, 10, 20, 30, 40, 50, 70, 120 GeV). 
At each energy point,100k events are simulated for each particle type. 
These samples follow a flat distribution in theta and phi over the 4$\pi$ solid angle. 

These samples are reconstructed with Arbor (version 3.3). To disentangle the lepton identification performance from the effect of PFA reconstruction and geometry defects, we select those events where only one charged particle is reconstructed. 
The total number of these events is recorded as $ N_{1 Particle}$, and the number of these events identified with correct particle types is recorded as $ N_{1 Particle, T}$. 
The performance of lepton identification is then expressed as a migration matrix in Table \ref{migrM}, 
its diagonal elements $\epsilon^{i}_{i}$ refer to the identification efficiencies (defined as $ N_{1 Particle, T}/N_{1 Particle}$), and the off diagonal element $P^{i}_{j}$ represent the probability of a type $i$ particle to be mis-identified as type $j$.

\begin{table}[htbp]
\centering
\caption{\label{migrM} Migration Matrix}
\smallskip
\begin{tabular}{ccccc}
\hline
  & $e^{-} like$ & $\mu ^{-} like$ & $\pi ^{+} like$ & undefined\\
\hline
 $e^{-}$ & $\epsilon ^{e} _{e}$ & $P^{e}_{\mu}$ & $P^{e}_{\pi}$ & $P^{e}_{und}$\\
 $\mu^{-}$ & $P ^{\mu} _{e}$ & $\epsilon^{\mu}_{\mu}$ & $P^{\mu}_{\pi}$  & $P^{\mu}_{und}$\\
  $\pi^{+}$ & $P ^{\pi} _{e}$ & $P^{\pi}_{\mu}$ & $\epsilon^{\pi}_{\pi}$ & $P^{\pi}_{und}$\\
\hline
\end{tabular}
\end{table}

\section{Discriminant variables and the output likelihoods}

LICH takes individual reconstructed charged particles as input, extracts 24 discriminant variables for the lepton identification, and calculates the corresponding likelihood to be an electron or a muon. 
These discriminant variables can be characterized into five different classes: 

\begin{itemize}
\item {\bf dE/dx}
 
For a track in the TPC, the distribution of energy loss per unit distance follows a Landau distribution. 
The dE/dx estimator used here is the average of this value but after cutting tails at the two edges of the Landau distribution (first 7\% and last 30\%). 
The dE/dx has a strong discriminant power to distinguish electron tracks from others at low energy (under 10 GeV) (Figure \ref{dedx}).

\begin{figure}[htbp]
\centering\includegraphics[%
  width=.48 \textwidth, clip, trim=0 30 0 10
  ]{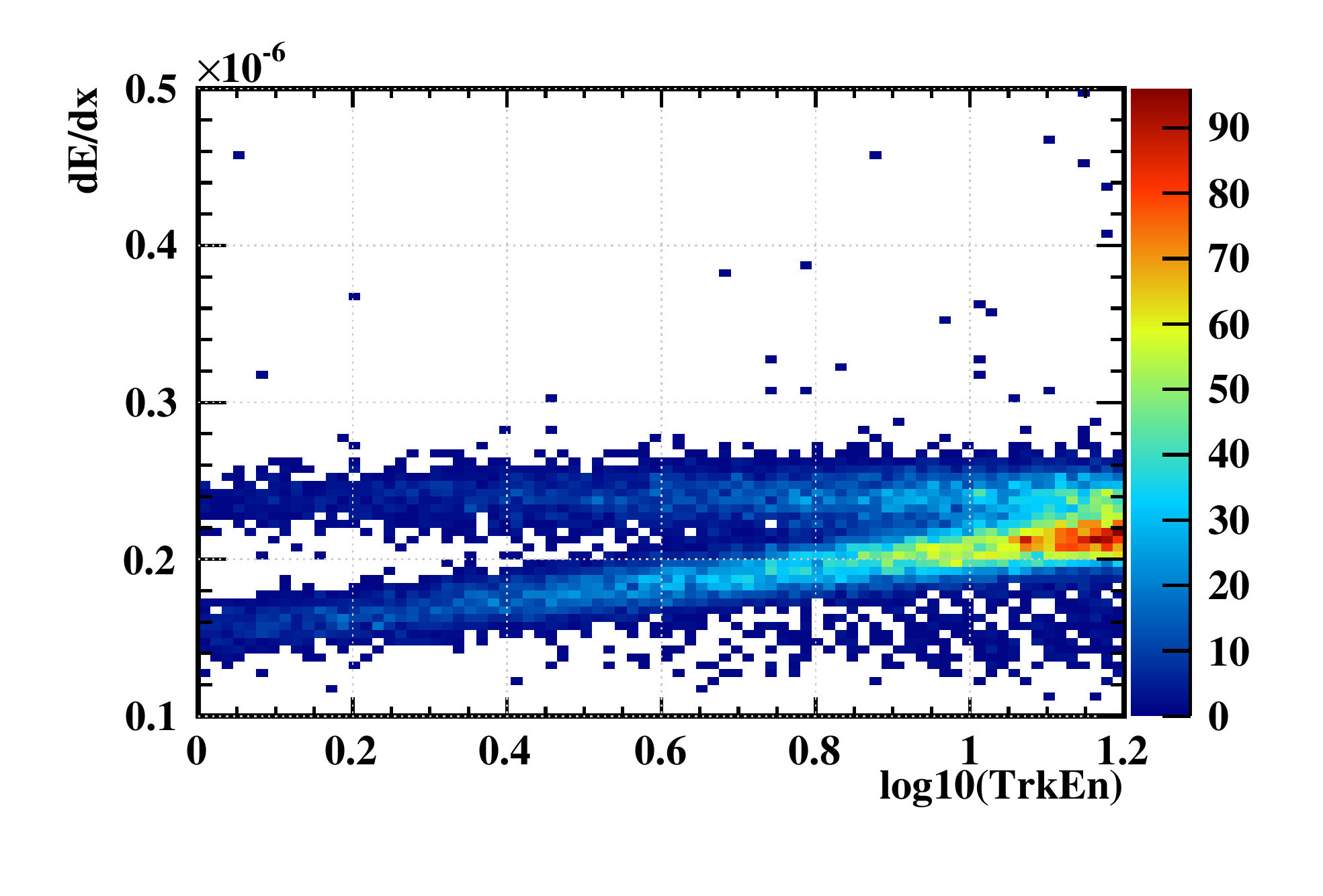} 
\caption{dE/dx for $e^{-}$, $\mu^{-}$ and $\pi ^{+}$, for electrons it is stable around $2.4 \times 10^{-7}$, for muon and pion it is smaller at energy lower than 10 GeV and after that they start mixing with electron}
\label{dedx}
\end{figure}

\vspace{3 mm}

\item { \bf Fractal Dimension}

The fractal dimension (FD) of a shower is used to describe the self-similar behavior of shower spatial configurations, following the original definition in \cite{FDmanqi}, the fractal dimension is directly linked to the compactness of the particle shower.

At a fixed energy, the EM showers are much more compact than the muon or hadron shower, leading to a large FD. 
The muon shower usually takes the configuration of a 1-dimensional MIP(Minimum Ionizing Particle) track, therefore has a FD close to zero. 
The FD of the hadronic shower usually lays between the EM and MIP tracks, since it contains both EM and MIP components. 
A typical distribution of FD for 40 GeV showers is presented in Figure \ref{ecalfd},

For any calorimeter cluster, LICH calculates 5 different FD values: from its ECAL hits, HCAL hits, hits in 10 or 20 first layers of ECAL, and all the calorimeter hits.

\begin{figure}[htbp]
\centering\includegraphics[%
  width=.4 \textwidth, clip,trim=0 0 0 20
  ]{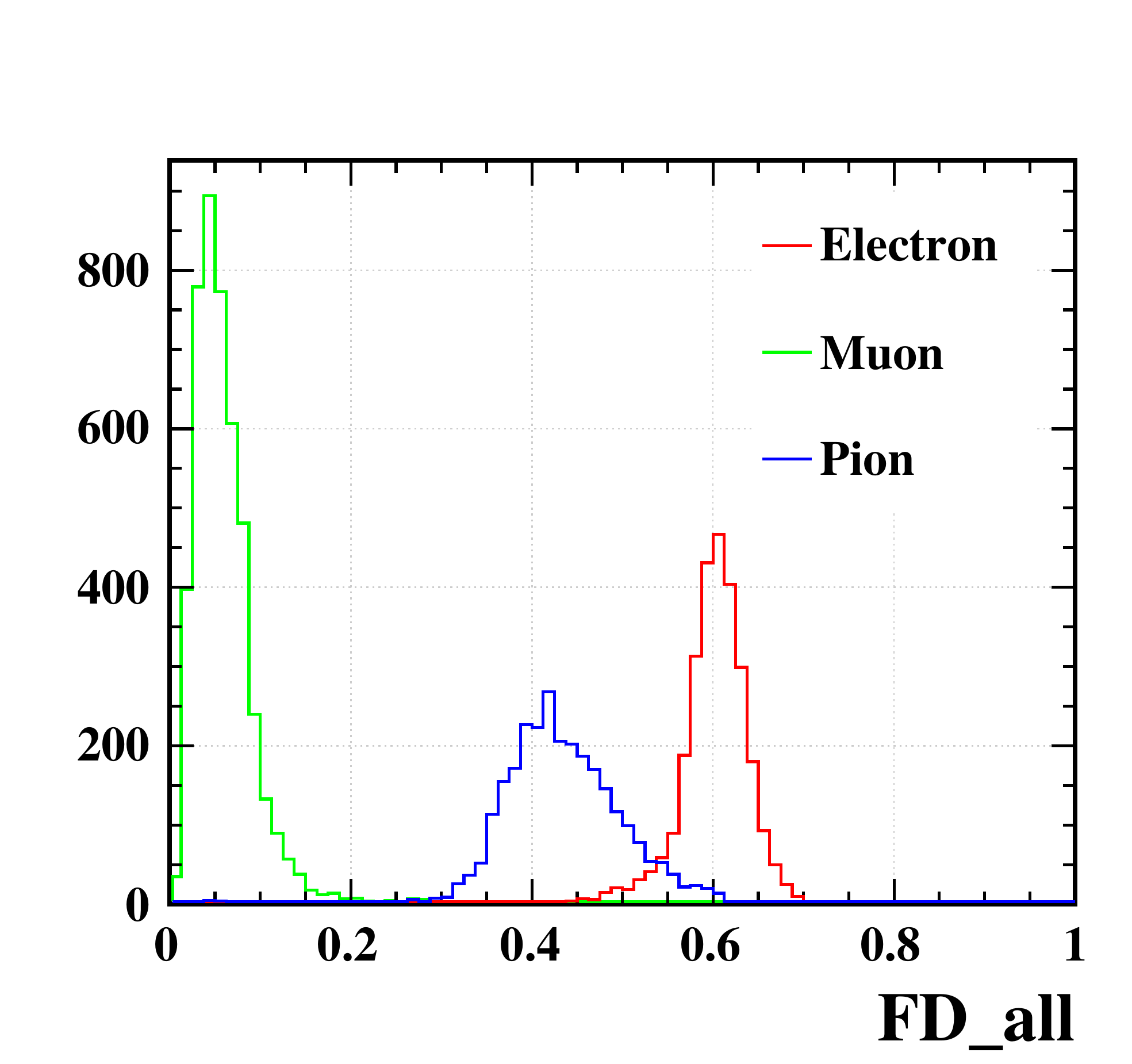} 
\caption{Fractal dimension using both ECAL and HCAL for $e^{-}$, $\mu^{-}$ and $\pi ^{+}$ at 40 GeV}
\label{ecalfd}
\end{figure}

\vspace{3 mm}

\item {\bf Energy Distribution} 

LICH builds variables out of the shower energy information, including the proportion of energy deposited in the first 10 layers in ECAL to the entire ECAL, or the energy deposited in a cylinder around the incident direction with a radius of 1 and $1.5$ Moliere radius. 

\vspace{3 mm}

\item {\bf Hits Information}

Hits information refers to the number of hits in ECAL and HCAL and some other information obtained from hits, such as the number of ECAL (HCAL) layers hit by the shower, number of hits in the first 10 layers of ECAL.

\vspace{3 mm}

\item {\bf Shower Shape, Spatial Information} 

The spatial variables include the maximum distance between a hit and the extrapolated track, the maximum distance and average distance between shower hits and the axis of the shower (defined by the innermost point and the center of gravity of the shower), the depth (perpendicular to the detector layers) of the center of gravity, and the depth of the shower defined as the depth between the innermost hit and the outermost hit. 

\end{itemize}

\begin{figure}[htbp]
\centering\includegraphics[%
  width=.48 \textwidth, clip, trim=0 60 0 20  
  ]{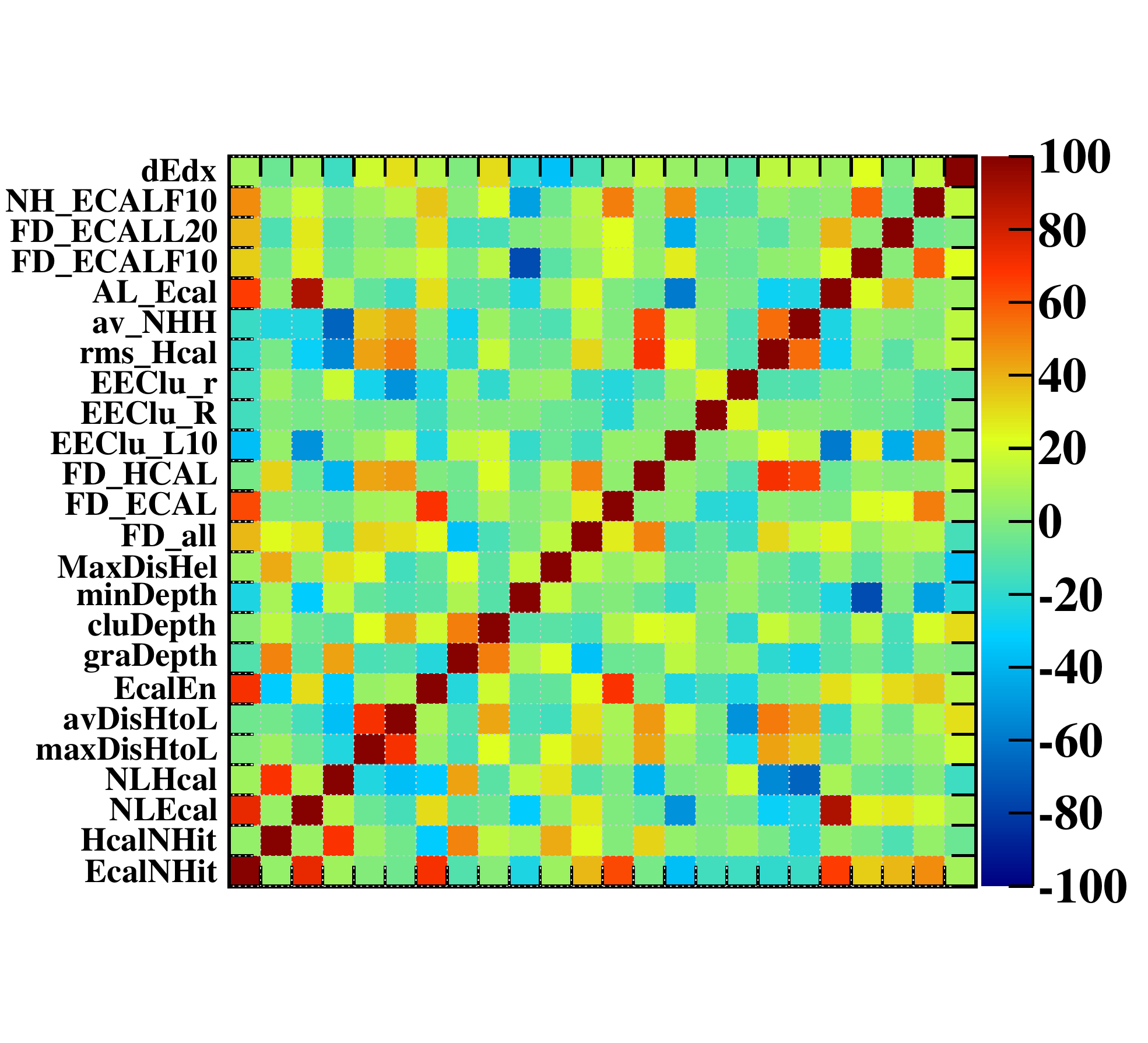} 
\caption{The correlation matrix of all the variables}
\label{CM}
\end{figure}

The correlation of those variables at energy 40 GeV are summarized in Figure \ref{CM}, the definitions of all the variables are listed in \ref{app}. It is clear that the dE/dx, measured from tracks, does not correlate with any other variables which are measured from calorimeters. Some of the variables are highly correlated, such as FD\_ECAL (FD calculated from ECAL hits) and EcalNHit (number of ECAL hits). However all these variables are kept because their correlations change with energy and polar angle.

\begin{figure}[htbp]
\centering\includegraphics[%
  width=.48 \textwidth, clip,trim=0 0 0 0
  ]{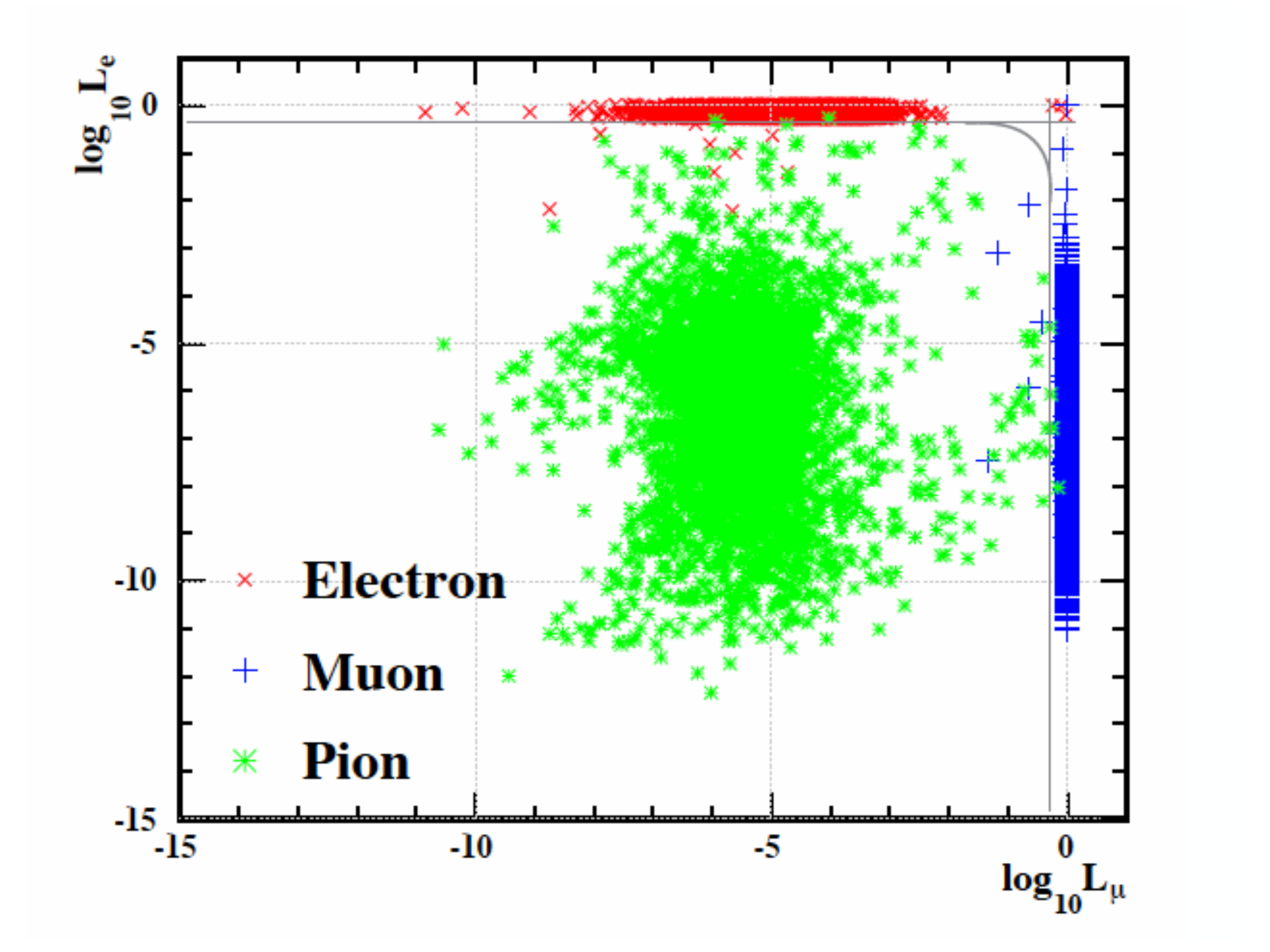} 
\caption{The e-likeliness and $\mu$-likeness of  $e^{-}$, $\mu^{-}$ and $\pi ^{+}$ at 40 GeV, grey lines are the cuts for different catalogs in next section}
\label{2DBDTG}
\end{figure}

LICH uses TMVA\cite{TMVA} methods to summarize these input variables into two likelihoods, corresponding to electrons and muons. 
Multiple TMVA methods have been tested and the Boosted Decision Trees with Gradient boosting (BDTG) method is chosen for its better performance. 
The e-likeness ($L_{e}$) and $\mu$-likeness ($L_{\mu}$) for different particles in a 40 GeV sample are shown in Figure \ref{2DBDTG}.

\section{Performance on single particle events}

The phase space spanned by the lepton-likelihoods ($L_{e}$ and $L_{\mu}$) can be separated into different domains, corresponding to different catalogs of particles. 
The domains for particles of different types can be adjusted according to physics requirements. 
In this paper, we demonstrate the lepton identification performance on single particle samples using the following catalogs:

\begin{itemize}

\item  Muon: $L_{\mu}$ > 0.5

\item Electron: $L_{e}$ > 0.5

\item Pion: 1-($L_{\mu}$+$L_e$)> 0.5

\item Undefined: $L_{\mu}$ <   0.5 \& $L_{e}$ <  0.5 \&  1-($L_{\mu}$+$L_{e}$) <  0.5

\end{itemize}

The probabilities of undefined particles are very low (<10$^{-3}$) at single particle samples with the above catalog. 

Since the distribution of these variables depends on the polar angle of the initial particle ($\theta$), the TMVA is trained independently on four subsets:
 
 \begin{itemize}
 
\item {\bf barrel 1}: middle of barrel ($| \cos \theta |< 0.3$),
\item {\bf barrel 2}: edge of barrel ($0.3 < |\cos \theta| < 0.7$), 
\item {\bf overlap}: overlap region of barrel and endcap ($0.7 <$ $|\cos \theta|$ $< 0.8$),  
\item {\bf endcap}: ($0.8 < |\cos \theta|$ < 0.98). 

\end{itemize}
 
 Take the sample of 40 GeV charged particle as an example, the migration matrix is shown in Table \ref{migrM}. 
 Comparing this table to the result of ALEPH for energetic taus\cite{ALEPH}, the efficiencies are improved, and the mis-identification rates from hadrons to leptons are significantly reduced.

\begin{table}[htbp]
\centering
\caption{\label{migrM} Migration Matrix at 40 GeV (\%)}
\smallskip
\begin{tabular}{lrrrr}
\hline
 Type & $e^{-}  like$ & $\mu ^{-}  like$ & $\pi ^{+}  like $\\
\hline 
 $e^{-}$ & $99.71 \pm 0.08$ & $<0.07$ & $0.21 \pm 0.07$\\
 $\mu^{-}$ & $<0.07$ & $99.87 \pm 0.08$ & $0.05 \pm 0.05$\\
  $\pi^{+}$ & $0.14 \pm 0.05$ & $0.35 \pm 0.08$ & $99.26 \pm 0.12$\\
\hline 
\end{tabular}
\end{table}

\begin{figure}[htbp]
\centering
\includegraphics[%
width=.48 \textwidth, clip, 
  ]{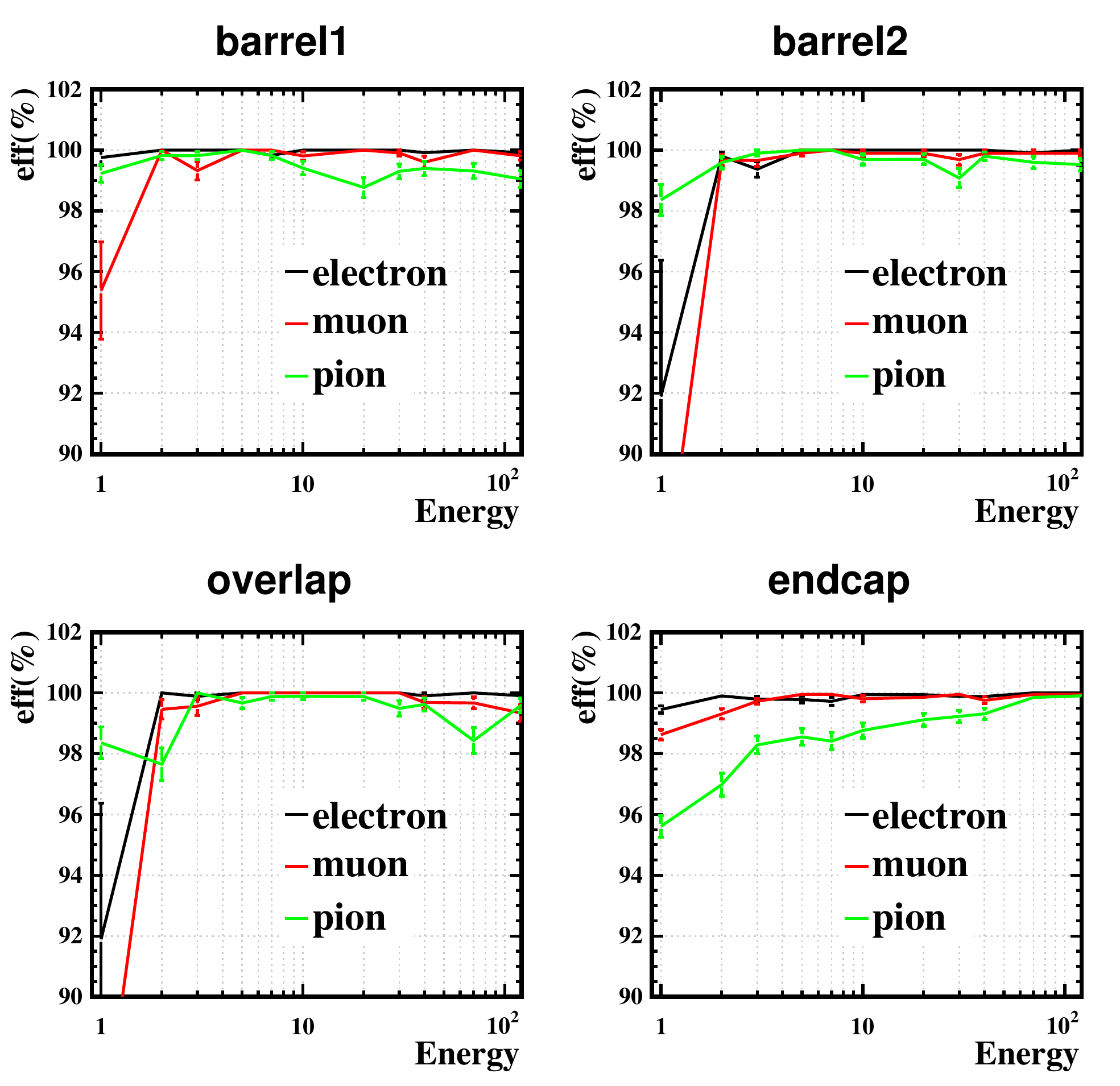} 
 \caption{The efficiency of lepton identification for $e^{-}$, $\mu^{-}$ and $\pi ^{+}$ as function of particle energy in the four regions}
\label{pideff}
\end{figure}

The lepton identification efficiencies (diagonal terms of the migration matrix) at different energies are presented in Figure \ref{pideff} for the different regions.  
The identification efficiencies saturate at 99.9\% for particles with energy higher than 2 GeV. 
For those with energy lower than 2 GeV, the performance drops significantly, especially in {\bf barrel2} and {\bf overlap} regions.
For the overlap region, the complex geometry limits the performance; while for the {\bf barrel2} region, charged particles with Pt < 0.97 GeV cannot reach the barrel, they will eventually hit the endcaps at large incident angle, hence their signal is more difficult to catalog.  

\begin{figure}[htbp]
   \centering\includegraphics[%
  width=.35 \textwidth, clip, trim=0 0 0 10
  ]{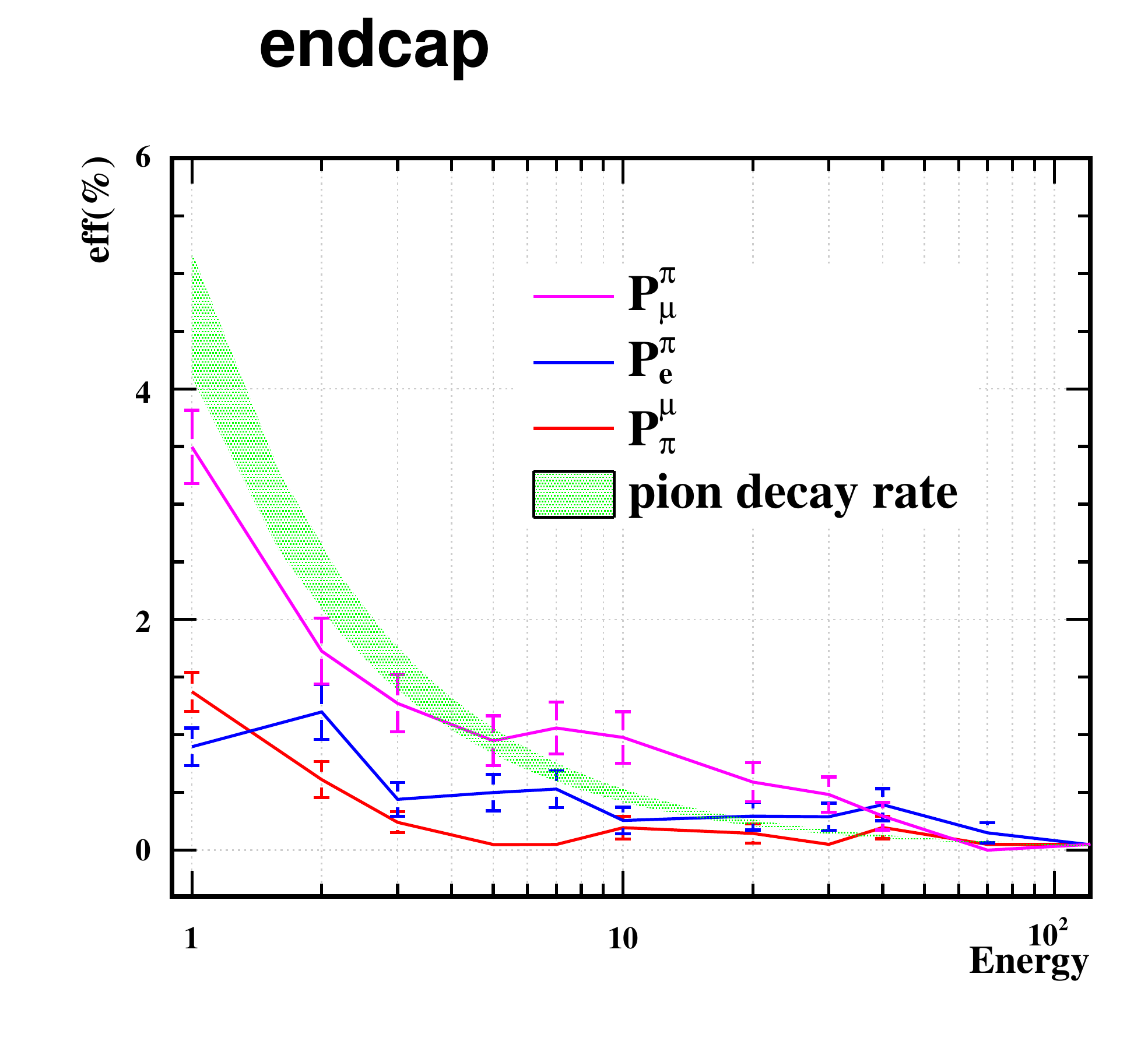} 
\caption{The mis-identification rates of lepton identification for $\mu$ and $\pi$ in $\sim$ 5000 events for the endcap region; Pion decay rate band (to account for the polar angle spread) is indicated for comparison}
\label{misID}
\end{figure}

Concerning the off-diagonal terms of the migration matrix, the chances of electrons to be mis-identified as muons and pions are negligible ($P^{e}_{\mu}, P^{e}_{\pi}<10^{-3}$), the crosstalk rate $P^{\mu}_{e}$ is observed at even lower level. 
However, the chances of pions to be mis-identified as leptons ($P^{\pi}_{e}$, $P^{\pi}_{\mu}$) are of the order of 1\% and are energy dependent.
In fact, these mis-identifications are mainly induced by the irreducible physics effects: pion decay and $\pi^{0}$ generation via $\pi$-nucleon collision. 
Meanwhile, the muons also have a small chance to be mis-identified as pions at energy smaller than 2 GeV. 
Figure \ref{misID} shows the significant crosstalk items ($P^{\pi}_{e}$, $P^{\pi}_{\mu} $and $P^{\mu}_{\pi}$) as a function of the particle energy in the endcap region.
The green shaded band indicates the probability of pion decay before reaching the calorimeter, which is roughly comparable with $P^{\pi}_{\mu}$.


\section{Lepton identification performance on single particle events for different geometries}

The power consumption and electronic cost of the calorimeter system scale with the number of readout channels.
It's important to evaluate the physics performance for different calorimeter granularities, at which the LICH performance is analyzed. 

The performance is scanned over certain ranges of the following parameters: 
\begin{itemize}
\item the number of layers in ECAL, taking the value of 20, 26, 30;
\item the number of layers in HCAL: 20, 30, 40, 48;
\item the ECAL cell size = 5$\times$5 mm$^{2}$, 10$\times$10 mm$^{2}$, 20$\times$20 mm$^{2}$, 40$\times$40 mm$^{2}$
\item HCAL cell size = 10$\times$10 mm$^{2}$, 20$\times$20 mm$^{2}$, 40$\times$40 mm$^{2}$, 60$\times$60 mm$^{2}$, 80$\times$80 mm$^{2}$
\end{itemize}

\begin{figure}[htbp]
\centering\includegraphics[%
  width=.48 \textwidth, clip,
    ]{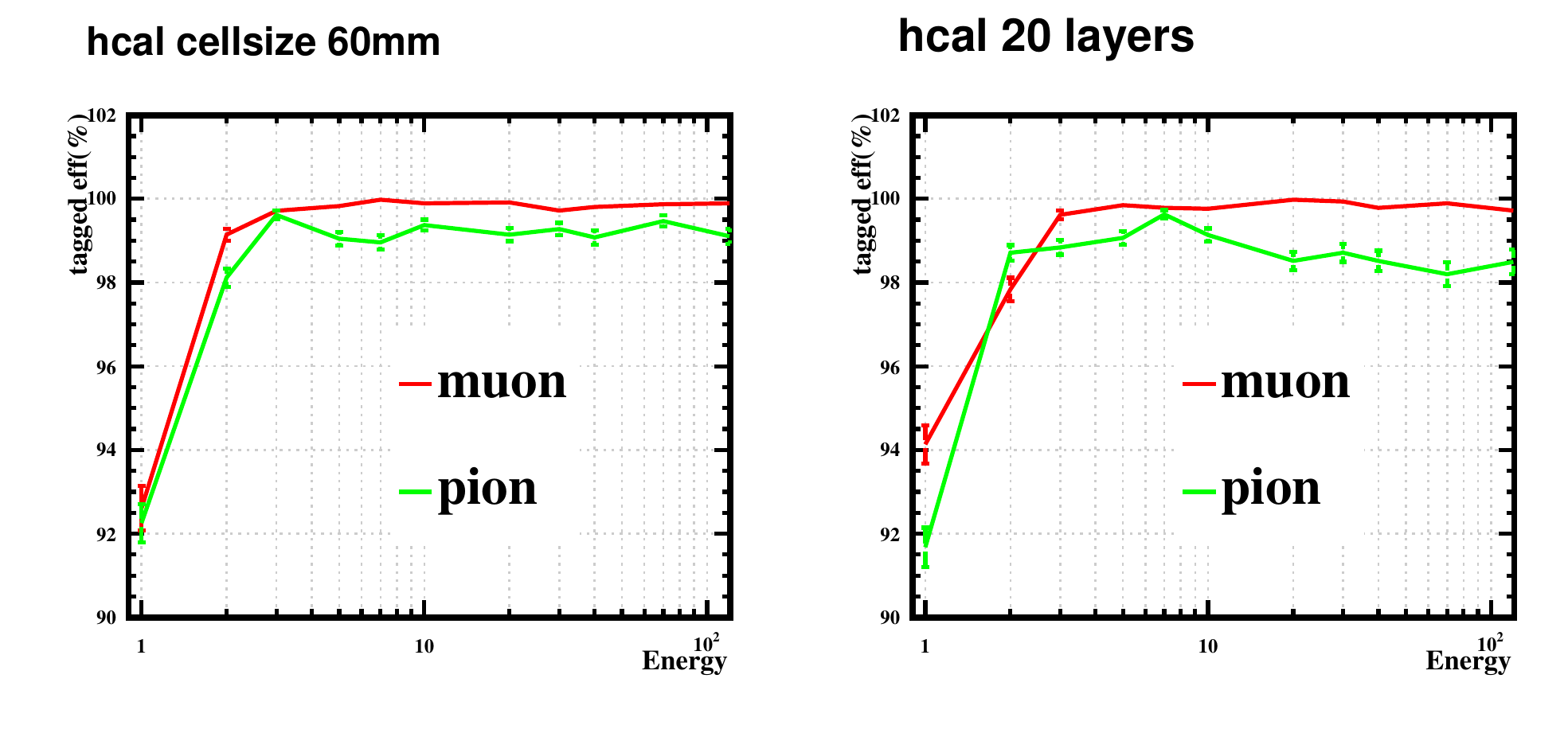} 
\caption{The efficiency of lepton identification for two different geometries}
\label{NSHeff}
\end{figure}

In general, the lepton identification performance is extremely stable over the scanned parameter space. 
Only for HCAL cell size larger than 60$\times$60 mm$^{2}$ or HCAL layer number less than 20, marginal performance degradation is observed:
the efficiency of identifying muons degrades by 1-2\% for low energy particles (E $\leq$ 2 GeV), and the identification efficiency of pion degrades slightly over the full energy range, see Figure \ref{NSHeff}.
  
\section{Performance on physics events}

\begin{figure}[htbp]
\centering
\includegraphics[%
  width=.2 \textwidth,clip
  ]{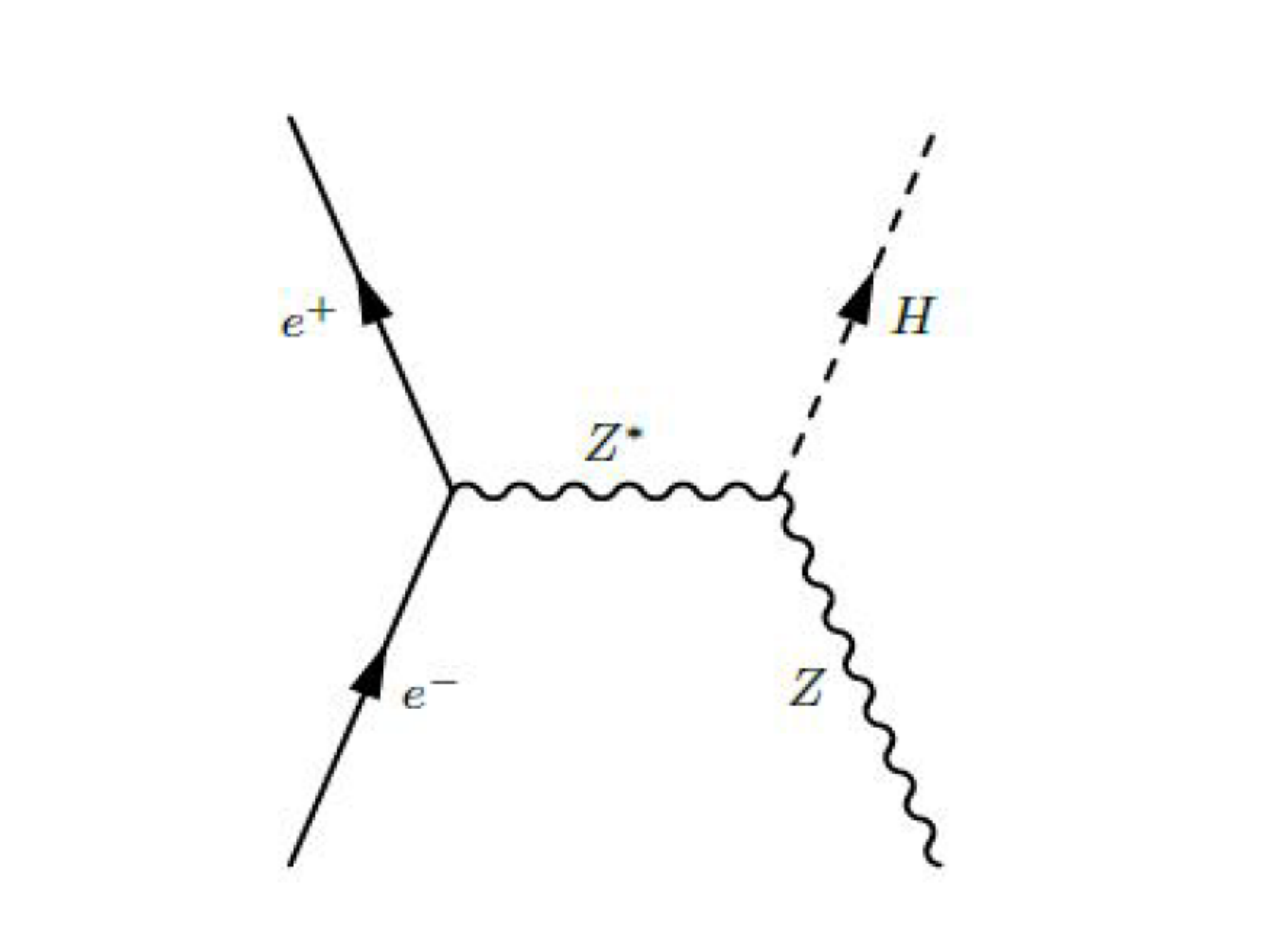} 
  \includegraphics[%
  width=.2 \textwidth,clip
  ]{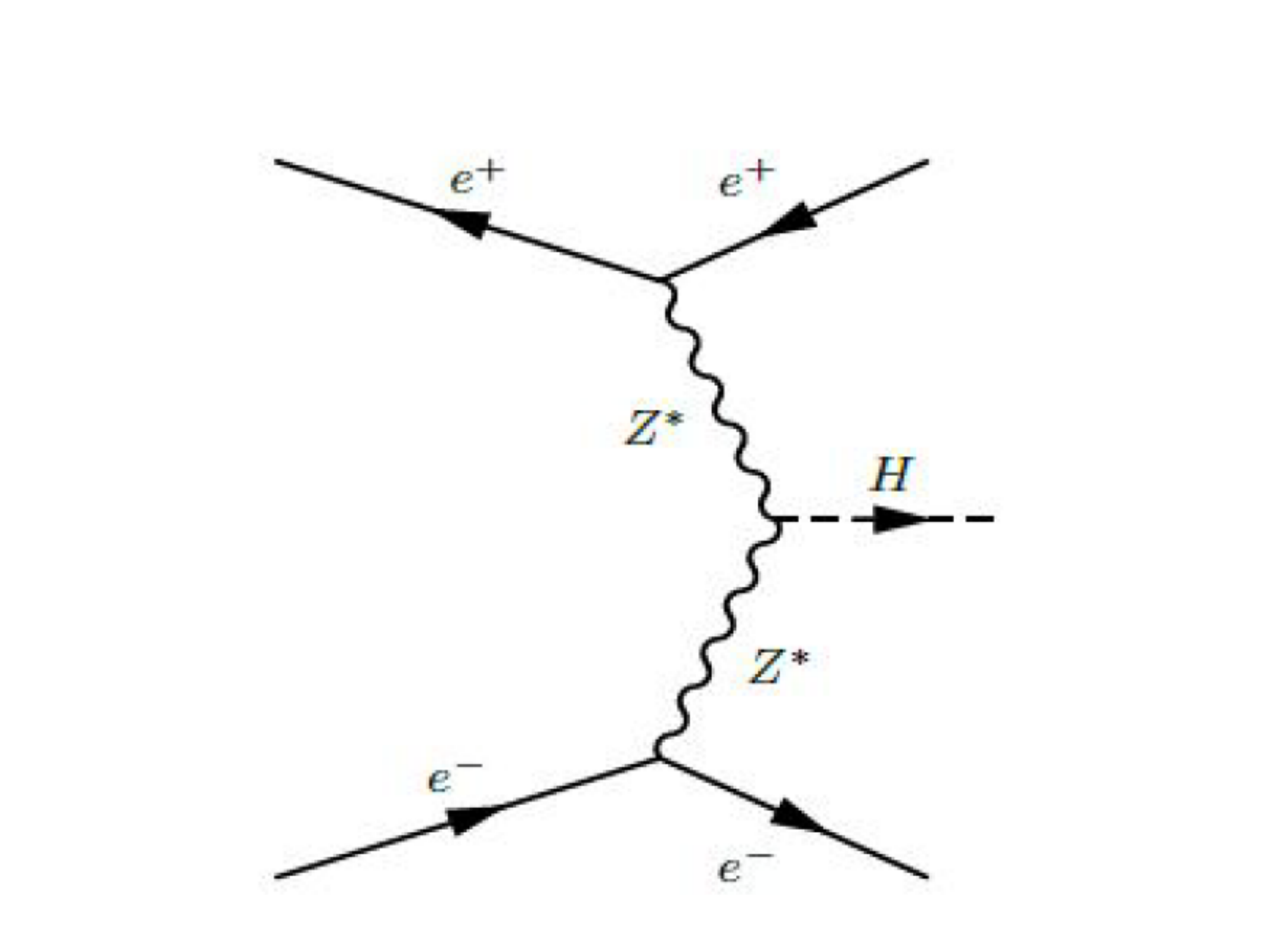} 
\caption{Feynman diagrams of major Higgs production with leptons at CEPC: the Higgsstrahlung and ZZ fusion processes.}
\label{FeymannDiag}
\end{figure}

The Higgs boson is mainly generated through the Higgsstrahlung process (ZH) and more marginally through vector boson fusion processes at electron-positron Higgs factories. 
A significant part of the Higgs bosons will be generated together with a pair of leptons (electrons and muons). 
These leptons are generated from the Z boson decay of the ZH process.
For the electrons, they can also be generated together with Higgs boson in the Z boson fusions events, see Figure \ref{FeymannDiag}.
At the CEPC, $3.6\times10^{4}$ $\mu\mu$H events and $3.9\times10^{4}$ eeH events are expected at an integrated luminosity of 5 ab$^{-1}$. In these events, the particles are rather isolated.
 
\begin{figure}[htbp]
\centering\includegraphics[%
  width=.22 \textwidth,clip
  ]{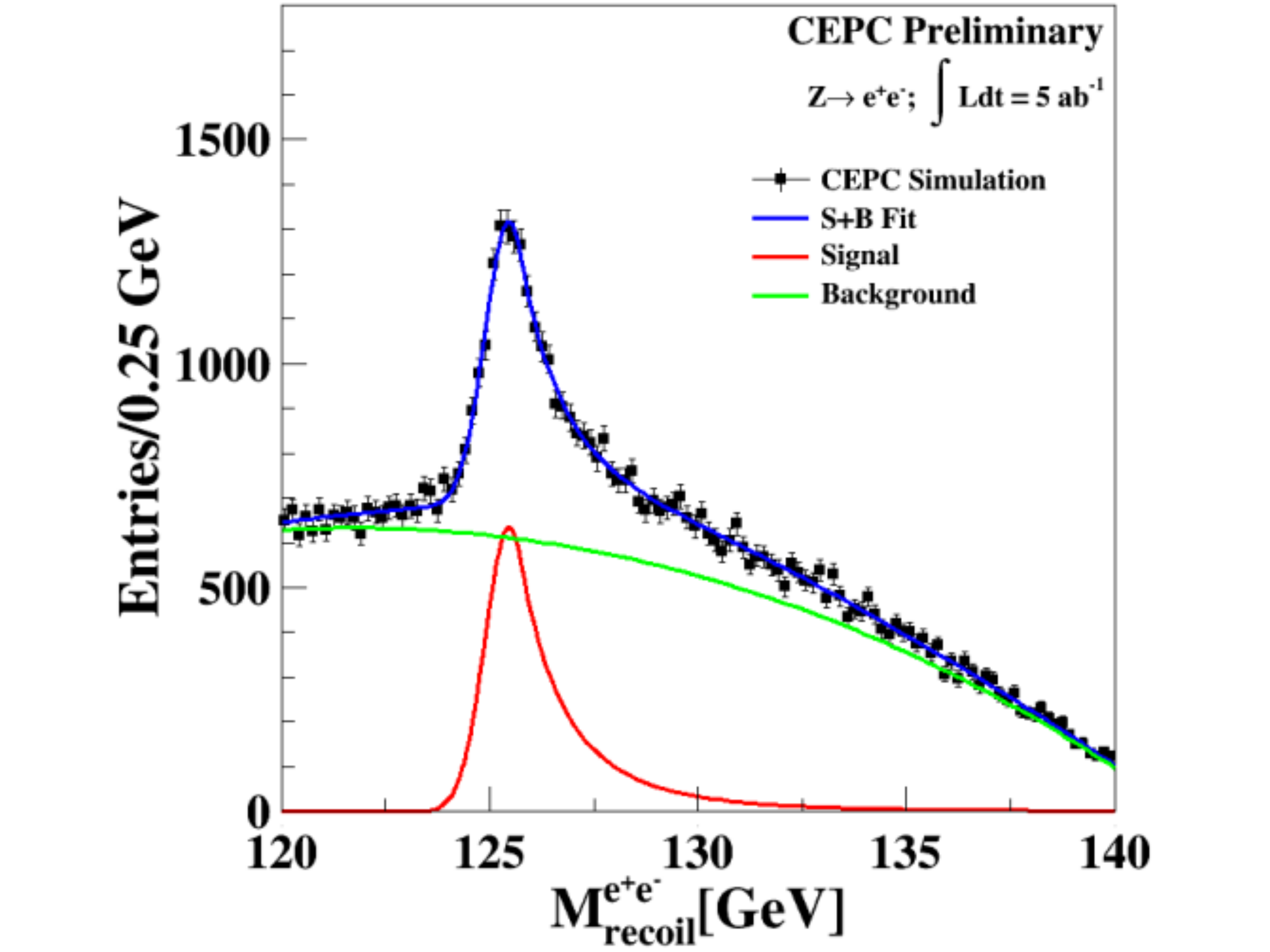} 
  \includegraphics[%
  width=.22 \textwidth,clip
  ]{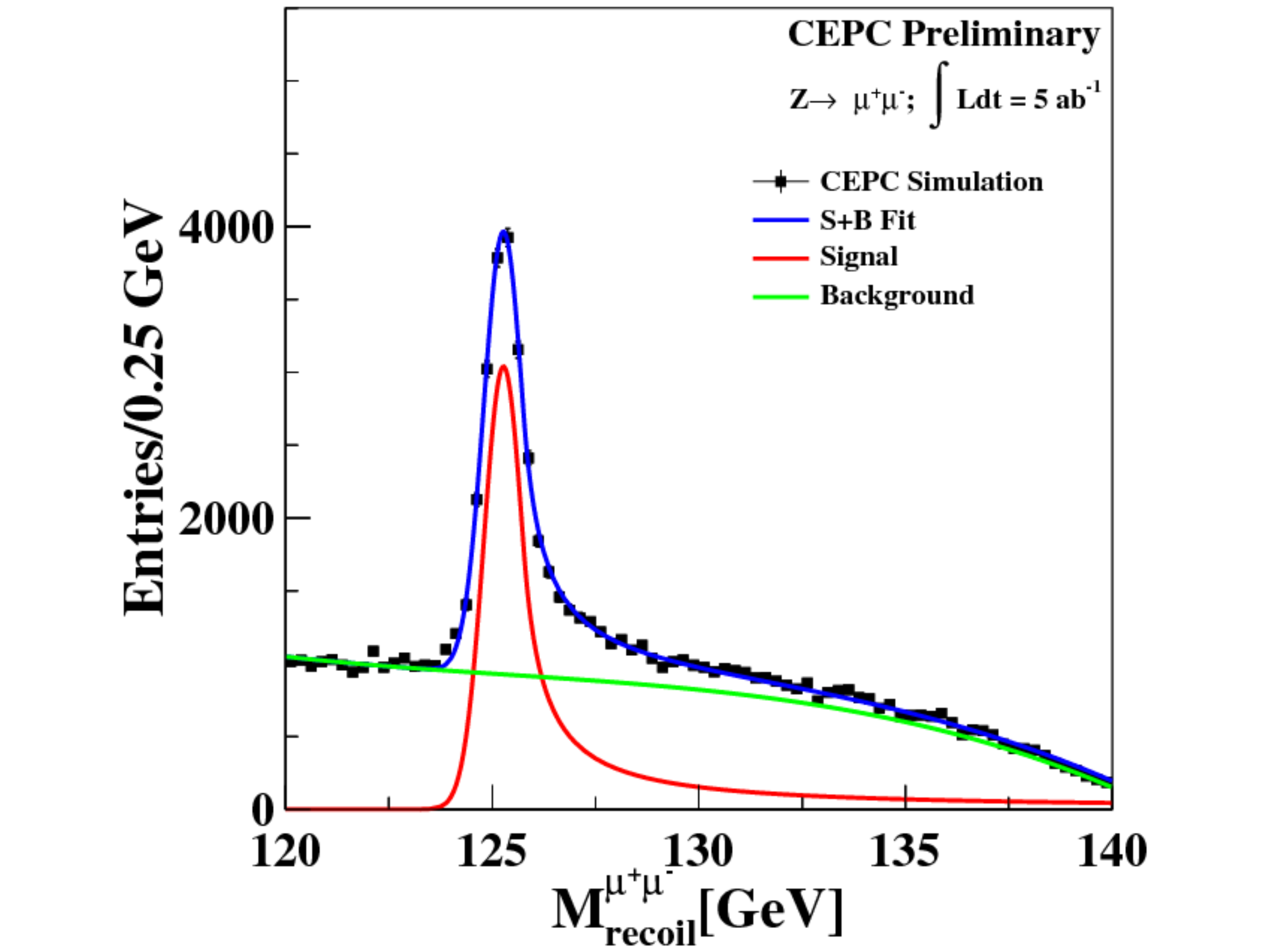} 
\caption{The recoil mass spectrum of ee/$\mu\mu$, low energy peak in eeH corresponds to the Z fusion events}
\label{recoilM}
\end{figure}

The eeH and $\mu\mu$H events provide an excellent access to the model-independent measurement to the Higgs boson using the recoil mass method \cite{mumuh}. 
The recoil mass spectrum of eeH and $\mu\mu$H events is shown in Figure \ref{recoilM}, which exhibits a high energy tail induced by the radiation effects (ISR, FSR, bremsstrahlung, beamstrahlung, etc), while in CEPC the beamstrahlung effect is negligible.
The bremsstrahlung effects for the muons are significantly smaller than that for the electrons, therefore, it has a higher maximum and a smaller tail.

\begin{figure}[htbp]
\centering\includegraphics[%
  width=.23 \textwidth,clip,trim=180 160 180 160
  ]{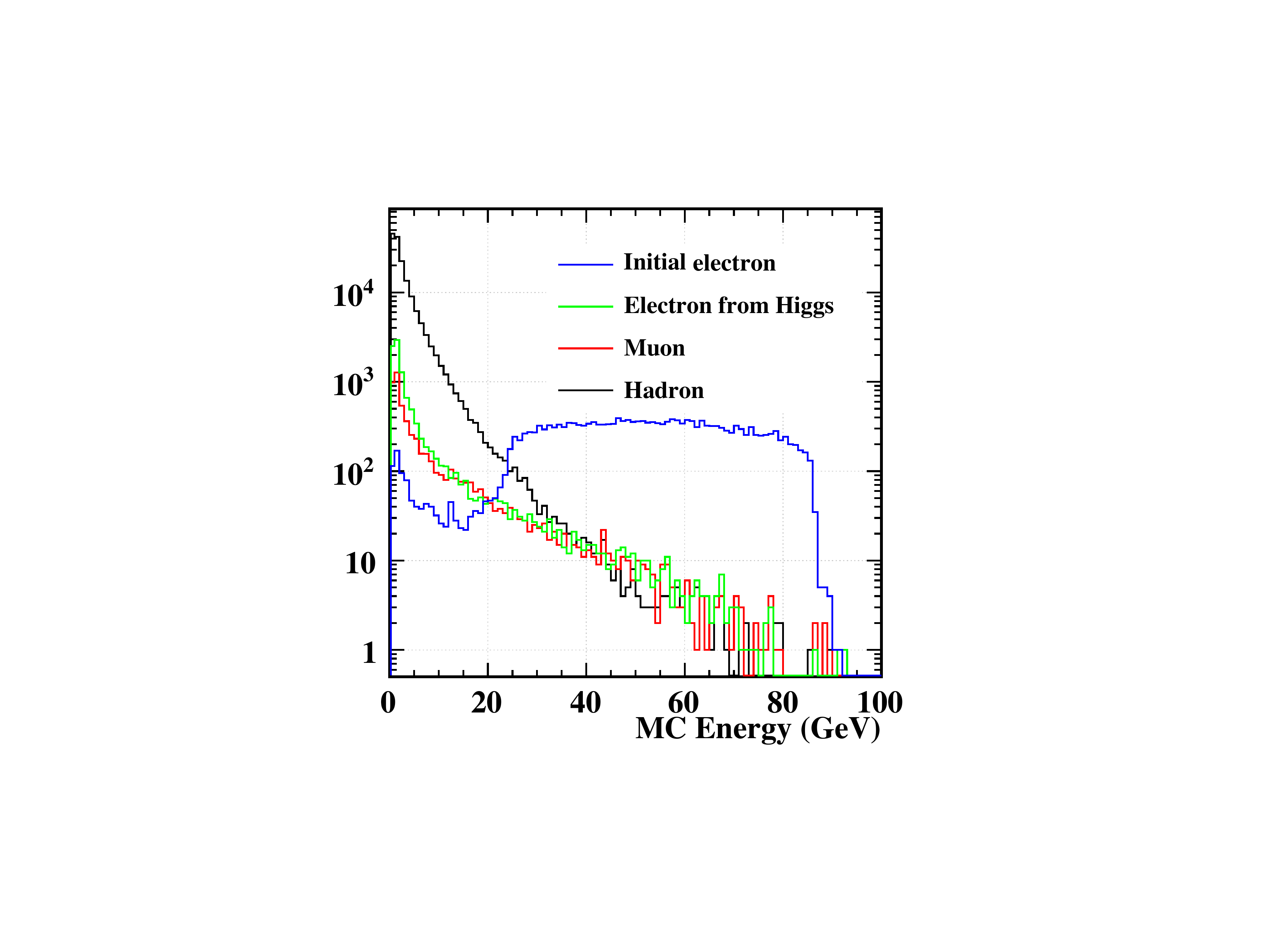}
  \includegraphics[%
  width=.23 \textwidth,clip,trim=180 160 180 160
  ]{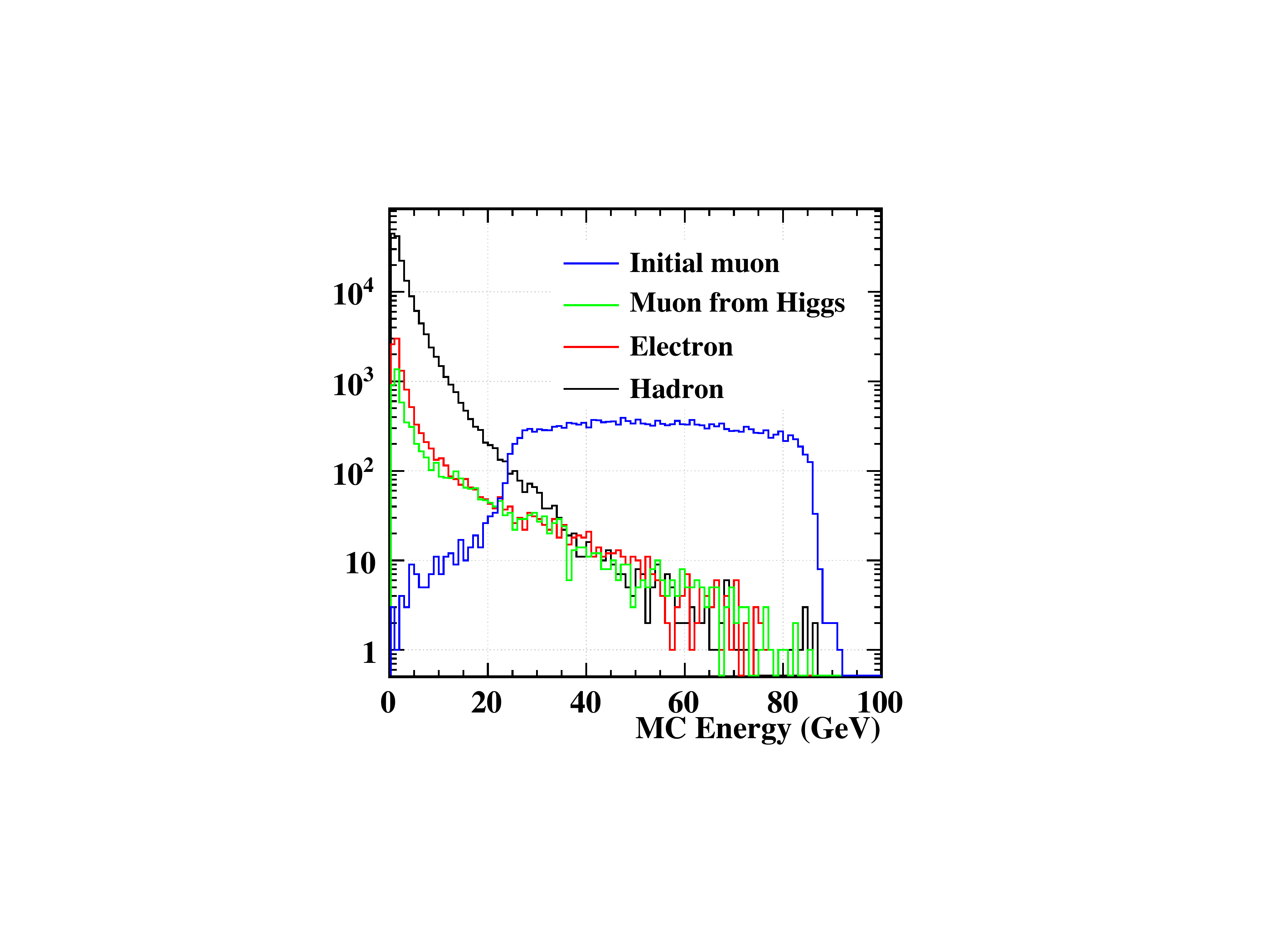}
  \caption{Energy Spectrum of charged particles in eeH event at 250 GeV center of mass energy}
  
  \label{EnSp}
  \end{figure}

Figure \ref{EnSp} shows the energy spectrum for all the reconstructed charged particles in 10k eeH/$\mu\mu$H events. 
The leptons could be classified into 2 classes, the initial leptons (those generated together with the Higgs boson) and those generated from the Higgs boson decay cascade.
For the eeH events, the energy spectrum of the initial electron exhibits a small peak at low energy, corresponding to the Z fusion events.
The precise identification of these initial leptons is the key physics objective for the lepton identification performance of the detector. 
 
Since the lepton identification performance depends on the particle energy,  and most of the initial leptons have an energy higher than 20 GeV,
we focused on the performance study of lepton identification on these high energy particles at detectors with two different sets of calorimeter cell sizes. 

\begin{figure*}[htbp]

\centering
\includegraphics[%
  width=.68 \paperwidth,clip, trim=0 350 0 30
   ]{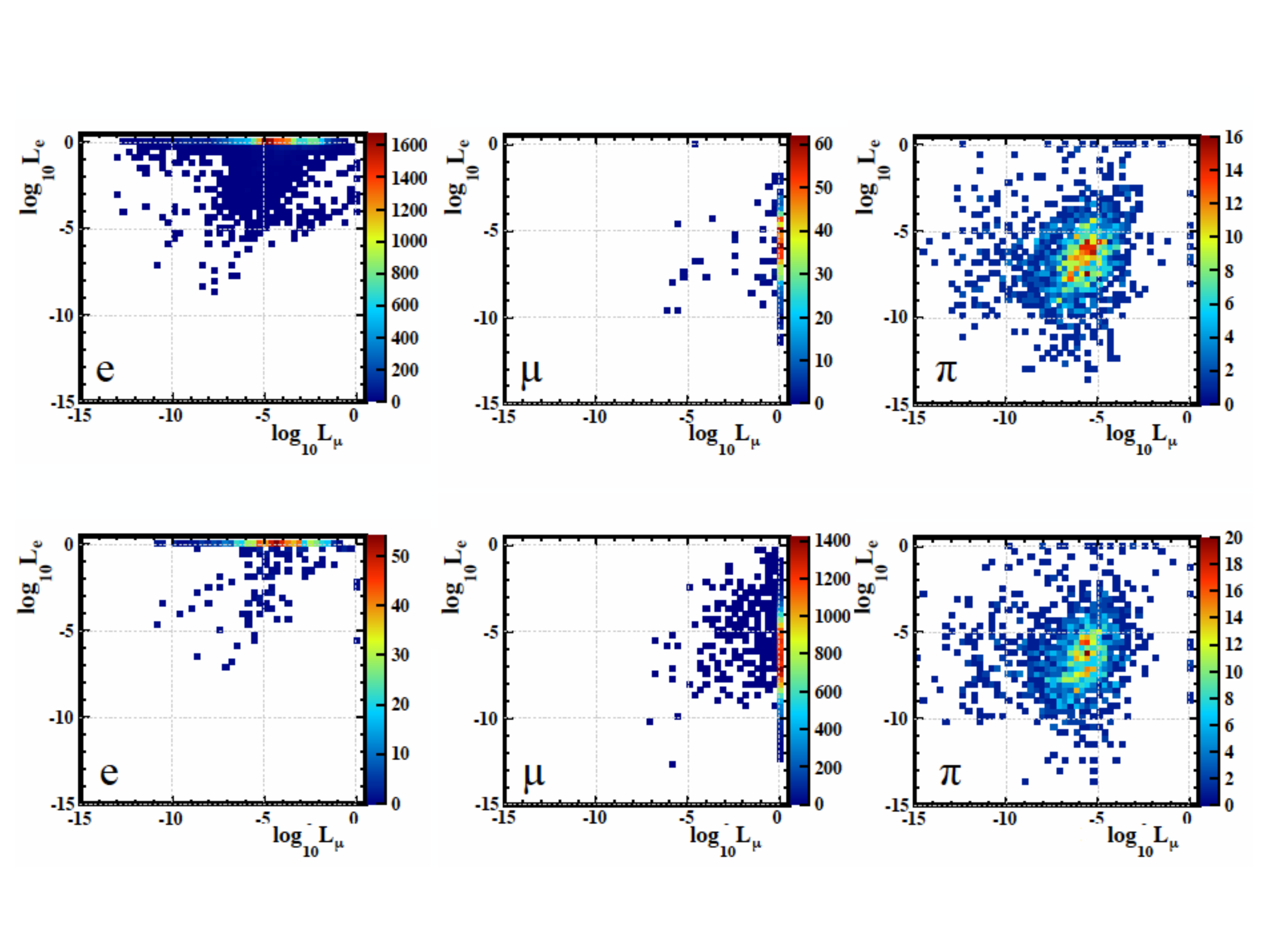} 
   \caption{e-likelihood and $\mu$-likelihood of charged particles with E>20 GeV in eeH event}
  \label{likeliness1}
\end{figure*}

\begin{figure*}[htbp]
\centering

\includegraphics[%
  width=.68 \paperwidth,clip,trim=0 20 0 400
  ]{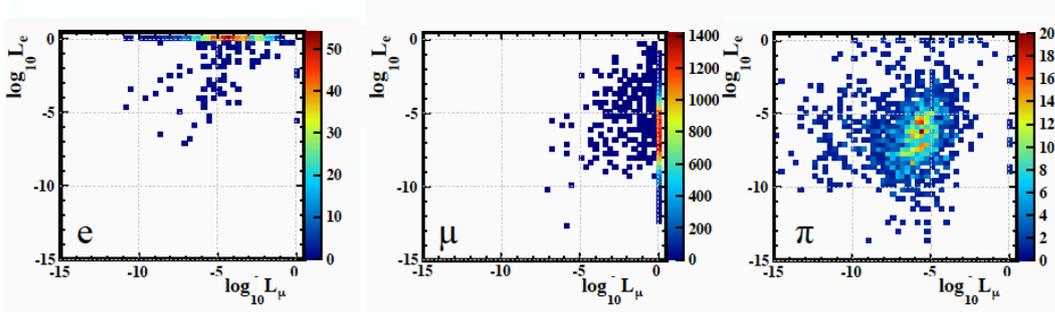} 
  \caption{e-likelihood and $\mu$-likelihood of charged particles with E>20 GeV in $\mu\mu$H event}
  \label{likeliness}
\end{figure*}

 The $\mu$-likeliness and e-likeliness of electrons, muons, and pions, for eeH events and $\mu\mu$H events are shown in Figure \ref{likeliness1} and Figure \ref{likeliness}. 
  Table \ref{table:hepeff} summarizes the definition of leptons and the corresponding performance at different conditions.   
The identification efficiencies for the initial leptons is degraded by 1-2\% with respect to the single particle case. 
This degradation is mainly caused by the shower overlap, and it's much more significant for electrons as electron showers are much wider than that of muon, leading to a larger chance of overlapping.  
The electrons in $\mu\mu$H events and vice versa, are generated in the Higgs decay. 
Their identification efficiency and purity still remains at a reasonable level.  
 For charged leptons with energy lower than 20 GeV, the performance degrades by about 10\% because of the high statistics of background and the cluster overlap. 
 The event identification efficiency, which is defined as the chance of successfully identifying both initial leptons, is presented in the last row of Table \ref{table:hepeff}. The event identification efficiencies is 
roughly the square of the identification efficiency of the initial leptons.  
Comparing the performance of both geometries, it is shown that when the number of readout channels is reduced by 4, the event reconstruction efficiency is degraded by 1.3\% and 1.7\%, for $\mu\mu$H and eeH events respectively. 
 
\begin{table*}
\centering
\caption{\label{table:hepeff} $\mu\mu$H/eeH  events lepton identification efficiency}
\smallskip
\begin{tabular*}{\textwidth}{@{\extracolsep{\fill}}lrrrrrl@{}}
\hline
 &  \multicolumn{2}{c}{Geom 1 (ECAL and HCAL Cell Size 10$\times$10 $mm^{2}$)} &  \multicolumn{2}{c}{Geom 2 (ECAL and HCAL Cell Size 20$\times$20 $mm^{2}$)}\\
\hline
& $\mu\mu$H & eeH &  $\mu\mu$H & eeH \\
 $\mu$ definition & $L_\mu$>0.1 & $L_\mu$>$ 0.1 $ & $L_\mu$>0.1 & $L_\mu$>0.1 \\
e definition & $L_e$>0.01 $L_\mu$<0.1 &  $L_e$>0.001 $L_\mu$<0.1 & $L_e$>0.01 $L_\mu$<0.1 & $L_e$>0.001 $L_\mu$<0.1 \\
$\varepsilon_{e}$ & $93.41\pm0.92$ & $\mathbf{98.64\pm0.08}$ & $91.60\pm1.02 $ & $\mathbf{97.89 \pm 0.11}$ \\
$\eta_{e}$ & $92.02\pm1.00$ & $ \mathbf{99.74\pm0.04}$ & $89.89 \pm 1.10 $ &  $\mathbf{99.67\pm0.04}$ \\
$\varepsilon_{\mu}$ & $\mathbf{99.54\pm0.05}$ & $95.53\pm0.76$ & $\mathbf{99.19\pm0.06}$ & $86.48\pm1.26$ \\
$\eta_{\mu}$ & $\mathbf{99.60\pm0.04}$ & $96.31\pm0.70$ & $\mathbf{99.83\pm0.03}$ & $95.38\pm0.81$ \\
\hline
$\varepsilon_{event}$ & $98.53\pm0.13$ & $97.06\pm0.19$ & $97.24\pm0.18$ & $95.40\pm0.24$ \\
\hline
\end{tabular*}
\end{table*}

\section{Conclusion}

The high granularity calorimeter is a promising technology for detectors in collider facilities of the High Energy Frontiers. It provides good separation between different final state particles,
which is essential for the PFA reconstructions. It also records the shower spatial development and energy profile to an unprecedented level of details, which can be used for the energy measurement and particle identifications.

To exploit the capability of lepton identification with high granularity calorimeters and also to provide a viable toolkit for the future Higgs factories, LICH, a TMVA based lepton identification package dedicated to high granular calorimeter, has been developed.
Using mostly the shower description variables extracted from the high granularity calorimeter and also the dE/dx information measured from tracker, LICH calculates the e-likeness and $\mu$-likeness for each individually reconstructed charged particle. Based on these output likelihoods, the leptons can be identified according to different physics requirement.

Applied to single particle samples simulated with the CEPC\_v1 detector geometry, the typical identification efficiency for electron and muon is higher than 99.5\% for energies higher than 2 GeV. For pions, the efficiency is reaching 98\%. These efficiencies are comparable to the performance reached by ALEPH, while the mis-identification rates are significantly improved. Ultimately, the performances are limited by the irreducible confusions, in the sense that the chance for muon to be mis-identified as electron and vice versa is negligible, the mis-identification of pion to muon is dominated by the pion decay.

The tested geometry uses a ultra-high granularity calorimeter: the cell size is 1 by 1 cm$^2$ and the layer number of ECAL/HCAL is 30/48. 
In order to reduce the total channel number, LICH is applied to a much more modest granularity, it is found that the lepton identification performance degrades only at particle energies lower than 2 GeV for an HCAL cell size bigger than 60$\times$60 mm$^{2}$ or with an HCAL layer number less than 20.

The lepton identification performance of LICH is also tested on the most important physics events at CEPC. In these events, multiple final state particles could be produced in a single collision, the particle identification performance will potentially be degraded by the overlap between nearby particles.
The lepton identification on eeH/$\mu\mu$H event at 250 GeV collision energy has been checked. The efficiency for a single lepton identification is consistent with the single particle results.
The efficiency of finding two leptons decreases by 1$\sim$2 \% when the cell size doubles, which means that the detector needs 2$\sim$4\% more statistics in the running. 
In eeH events, the performance degrades because the clustering algorithm still needs to be optimized.

To conclude, ultra-high granularity calorimeter designed for ILC provides excellent lepton identification ability, for operation close to ZH threshold. It may be a slight overkill for CEPC and a slightly reduced granularity can reach a better compromise. And LICH, the dedicated lepton identification for future e+e- Higgs factory, is prepared.

\begin{acknowledgements}

This study was supported by National Key Programme for S\&T Research and Development (Grant NO.: 2016YFA0400400), the Hundred Talent programs of Chinese Academy of Science No. Y3515540U1, and AIDA2020.

\end{acknowledgements}

\appendix
\section{Appendix section}\label{app}

List and meaning of variables used in the TMVA which are not mentioned in the text:

\begin{itemize}
\item NH\_ECALF10: Number of hits in the first 10 layers of ECAL
\item  FD\_ECALL20: FD calculated using hits in the last 20 layers of ECAL
\item  FD\_ECALF10: FD calculated using hits in the first 10 layers of ECAL
\item AL\_ECAL: Number of ECAL layer groups (each five layers forms a group) with hits
\item av\_NHH: Average number of hits in each HCAL layer groups (each five layers forms a group)
\item rms\_Hcal: The RMS of hits in each HCAL layer groups (each five layers forms a group)
\item EEClu\_r: Energy deposited in a cylinder around the incident direction with a radius of 1 Moliere radius
\item EEClu\_R: Energy deposited in a cylinder around the incident direction with a radius of 1.5 Moliere radius
\item EEClu\_L10: Energy deposited in the first 10 layers of ECAL
\item MaxDisHel: Maximum distance between a hit and the helix
\item minDepth: Depth of the inner most hit
\item cluDepth: Depth of the cluster position
\item graDepth: Depth of the cluster gravity center
\item EcalEn: Energy deposited in ECAL
\item avDisHtoL: Average distance between a hit to the axis from the inner most hit and the gravity center
\item maxDisHtoL: Maximum distance between a hit to the axis from the inner most hit and the gravity center
\item NLHcal: Number of HCAL layers with hits
\item NLEcal: Number of ECAL layers with hits
\item HcalNHit: Number of HCAL hits
\item EcalNHit: Number of ECAL hits

\end{itemize}


\begin{thebibliography}{9}

\bibitem{ILCTDR}
T. Behnke, J.E. Brau, P.N. Burrows, et al, The International Linear Collider Technical Design Report-Volume 4: Detectors[J]. arXiv preprint arXiv:1306.6329, 2013.

\bibitem{peskin}
M.E. Peskin, Physics goals of the linear collider[J]. arXiv preprint hep-ph/9910521, 1999.

\bibitem{atlas}
ATLAS collaboration, Physics at a High-Luminosity LHC with ATLAS[J]. arXiv preprint arXiv:1307.7292, 2013.

\bibitem{cms}
CMS collaboration, Projected Performance of an Upgraded CMS Detector at the LHC and HL-LHC: Contribution to the Snowmass Process[J]. arXiv preprint arXiv:1307.7135, 2013.

\bibitem{clic}
CLIC CDR, A multi-TeV linear collider based on CLIC technology: CLIC Conceptual Design Report[J]. edited by M. Aicheler, P. Burrows, M. Draper, T. Garvey, P. Lebrun, K. Peach, N. Phinney, H. Schmickler, D. Schulte and N. Toge, CERN-2012-007, 2012.

\bibitem{cepcprecdr}
M. Ahmad et al (The CEPC-SPPC Study Group), CEPC-SppC Preliminary Conceptual Design Report: Physics and Detector, http://cepc.ihep.ac.cn/preCDR/main preCDR.pdf, retrieved 4th May 2015

\bibitem{mumuh}
Z. Chen,Y. Yang, M. Ruan, et al, Study of Higgsstrahlung Cross Section and Higgs Mass Measurement Precisions with ZH ($ Z \rightarrow \mu^{+} \mu^{-} $) events at CEPC[J]. arXiv preprint arXiv:1601.05352, 2016.

\bibitem{cmsupg}
CMS collaboration, Technical proposal for the phase-II upgrade of the CMS detector[J]. CERN, CERN-LHCC-2015-010. LHCC-P-008, 2015.

\bibitem{pfa}
M.A. Thomson, Particle flow calorimetry and the PandoraPFA algorithm[J]. Nuclear Instruments and Methods in Physics Research Section A: Accelerators, Spectrometers, Detectors and Associated Equipment, 2009, 611(1): 25-40.

\bibitem{cmspfa}
F. Beaudette, The CMS Particle Flow Algorithm[J]. arXiv preprint arXiv:1401.8155, 2014.

\bibitem{jcpfa}
J.C. Brient, Improving the Jet Reconstruction with the Particle Flow Method; an Introduction[J]. arXiv preprint physics/0412149, 2004.

\bibitem{eepfa}
J.C. Brient, H. Videau, The calorimetry at the future e+ e-linear collider[J]. arXiv preprint hep-ex/0202004, 2002.

\bibitem{henripfa}
H. Videau, Energy flow or Particle flow-The technique of energy flow for pedestrians[C]//International Conference on Linear Colliders-LCWS04. Ecole Polytechnique Palaiseau, 2004: 105-120.


\bibitem{Arbor} 
M. Ruan, Arbor, a new approach of the Particle Flow Algorithm. 
arXiv:1403.4784 (2014).

\bibitem{ildloi}
T. Abe, ILD Concept Group-Linear Collider Collaboration. The International Large Detector: Letter of Intent, 2010[J]. arXiv preprint arXiv:1006.3396, 4(10).

\bibitem{FDmanqi} 
M. Ruan, D. Jeans, V. Boudry, J.C. Brient, \& H. Videau, (2014), Fractal Dimension of Particle Showers Measured in a Highly Granular Calorimeter, \emph{Physical review letters}, 112(1), 012001.





\bibitem{TMVA} A. Hoecker, P. Speckmayer,J. Stelzer , J. Therhaag, E. von Toerne, H. Voss, ... \& D. Dannheim (2007), TMVA-Toolkit for multivariate data analysis. 
arXiv preprint physics: 0703039.


\bibitem{ALEPH} Aleph Collaboration, Measurement of the Tau Polarisation at LEP[J]. arXiv preprint hep-ex/0104038, 2001.




\end{thebibliography}
\end{document}